\documentclass[article]{aa}
\usepackage{graphicx}
\usepackage{ae}
\usepackage{txfonts}
\usepackage{natbib}
\bibpunct{(}{)}{;}{a}{}{,}

\begin{document}

\title{Protostellar angular momentum evolution during gravoturbulent fragmentation }

\author{A.-K.\ Jappsen\and R.\ S.\ Klessen} \authorrunning{Jappsen \&
  Klessen} \titlerunning{Angular momentum evolution} \offprints{A.-K.\ 
  Jappsen, \email{akjappsen@aip.de}}

\institute{Astrophysikalisches Institut Potsdam, An der Sternwarte 16, D-14482 Potsdam, Germany}

\date{Received <date> / Accepted <date>}

\abstract{ Using hydrodynamic simulations we investigate the rotational
  properties and angular momentum evolution of prestellar and protostellar
  cores formed from gravoturbulent fragmentation of interstellar gas clouds.
  We find the specific angular momentum~$j$ of molecular cloud cores in
  the prestellar phase to be on average $\langle j \rangle =
  7\times10^{20}\,\mathrm{cm^2\,s^{-1}}$ in our
  models.  This is comparable to the observed
  values. A fraction of those cores is gravitationally unstable and goes
  into collapse to build up protostars and protostellar systems, which then have
  $\langle j \rangle = 8\times10^{19}\,\mathrm{cm^2\,s^{-1}}$. This is one order
  of magnitude lower than their parental cores and in agreement with
  observations of main-sequence binaries.
  The loss of specific angular momentum during collapse is mostly due to
  gravitational torques exerted by the ambient turbulent flow as well as by
  mutual protostellar encounters in a dense cluster environment.  Magnetic
  torques are not included in our models, these would lead to even larger
  angular momentum transport.
  
  The ratio of rotational to gravitational energy~$\beta$ in 
  cloud  cores that go into gravitational collapse turns out to be 
  similar to the observed values. We
  find, $\beta$ is roughly conserved during the main collapse phase. This
  leads to the correlation $j \propto M^{2/3}$, between specific angular
  momentum $j$ and core mass $M$.  Although the temporal evolution of the
  angular momentum of individual protostars or protostellar systems is complex
  and highly time variable, this correlation holds well in a statistical sense
  for a wide range of turbulent environmental parameters. In addition, high
  turbulent Mach numbers result in the formation of more numerous protostellar
  cores with, on average, lower mass. Therefore, models with larger Mach
  numbers result in cores with lower specific angular momentum.  We find,
  however, no dependence on the spatial scale of the turbulence.  Our models
  predict a close correlation between the angular momentum vectors of
  neighboring protostars during their initial accretion phase. Possible
  observational signatures are aligned disks and parallel outflows. The latter
  are indeed observed in some low-mass isolated Bok globules.
   
}
\maketitle
\section{Introduction}

Angular momentum plays a pivotal role in star formation.  The amount
of specific angular momentum determines whether a collapsing
protostellar core will form a single star or a binary or higher-order
multiple system. 

Stars are thought to form by gravoturbulent fragmentation in
interstellar clouds. The supersonic turbulence ubiquitously observed
in molecular gas generates strong density fluctuations with gravity
taking over in the densest and most massive regions.  Once gas clumps
become gravitationally unstable, collapse sets in and the central
density increases until a protostellar object forms and grows in mass
via accretion from the infalling envelope. Various aspects of the
relation between supersonic turbulence and star formation have been
discussed, e.g., by \citet{HUN82}, \citet{ELM93}, \citet{LAR95},
\citet{PAD95}, \citet{BAL99a,BAL99b,BAL03}, \citet{PAD99,PAD02},
\citet{VAZ00}, \citet*{KLE00b}, \citet*{HEI01}, \citet{KLE01c,KLE01a},
\citet{GAM03}, or \citet*{VAZ03}. In particular see the reviews by
\citet{LAR03} and \citet{MAC04}.

This dynamic picture of gravoturbulent star formation challenges the
so called ``standard theory'' where stars build up from the
``inside-out'' collapse of singular isothermal spheres, which are
generally assumed to result from the quasistatic contraction of
magnetically supported cloud cores due to ambipolar diffusion
\citep*{SHU77,SHU87}. This picture, however, has always
received strong criticism \citep[e.g.,][ for a critical
discussion]{WHI85,WHI96,NAK98}.  In particular, it seems strongly
biased toward the formation of single stars \citep{WHI96} which is in
contradiction to the observational fact that most (if not all) stars
form as members of binary or higher-order multiple systems \citep[see,
e.g., the reviews by][ and references therein]{BOD00,MAT00}.

Gravitational collapse in the astrophysical context always involves solving
the angular momentum problem.  It results from the blatant discrepancy between
the specific angular momentum observed in low-density gas on large scales and
the amount of rotation present after collapse \citep{SPI68,BOD95}.  
The source
of angular momentum on large scales lies in the differential rotation of the
galactic disk and, closely related to that, on intermediate to small scales it
results from the high degree of vorticity inextricably adherent to turbulent
flows.  
The typical specific angular momentum $j$ of molecular cloud material,
e.g., on scales of about 1$\,$pc is $j\approx 10^{23}\,$cm$^2$\,s$^{-1}$, while
on scales of cloud cores, say below $0.1\,$pc, it is of order of
$10^{21}\,$cm$^2$\,s$^{-1}$. A binary G star with a  orbital period of 3 days
has $j \approx 10^{19}\,$cm$^2$\,s$^{-1}$, while the spin of a typical T Tauri
star is a few $\times 10^{17}\,$cm$^2$\,s$^{-1}$. Our own Sun rotates only with
$j\approx 10^{15}\,$cm$^2$\,s$^{-1}$. That means, during the process of star
formation most of the initial angular momentum is removed from the collapsed
object.

The presence of magnetic fields, in principle, provides a viable mechanism for
locally reducing the angular momentum through magnetic braking. This was
treated approximately by \citet*{EBE60}, and later calculated accurately by
\citet{MOU79,MOU80}.  The criterion for effective braking is essentially that
the outgoing helical Alfv\'en waves from the rotating cloud have to couple to
the ambient medium over a volume that contains roughly the same mass as the
cloud itself. For the strong magnetic fields required by the standard theory
of star formation, the deceleration time can be less than the free-fall time,
leading to efficient transfer of angular momentum away from collapsing cores,
and thus, to the formation of single stars. Field strengths small enough to
allow for binary formation cannot provide support against collapse, thus pointing
toward a more dynamic picture of star formation as offered by gravoturbulent
fragmentation.

It is therefore a crucial test for any theory of star formation
whether it can produce the required angular momentum loss during
collapse while at the same time explain the high numbers of binaries
and multiple stellar systems observed \citep[e.g.,][]{DUQ91b,HAL03}.  In a
  semiempirical analysis of isolated binary star formation \citet{FIS04}
  presented the effects of turbulence in the initial state of the gas on
  binary orbital parameters. These properties were in agreement with
  observations if a significant loss of angular momentum was assumed. In the current investigation we focus on
models of non-magnetic, supersonically turbulent, self-gravitating
clouds and analyze the time evolution of angular momentum during
formation and subsequent collapse of protostellar cores. Our main
question is whether gravoturbulent fragmentation can solve or at least
ease the so called ``angular momentum'' problem without invoking the
presence of magnetic fields.

The structure of the paper is as follows. In Section~\ref{sec:models}
we introduce and discuss the numerical method to calculate the
dynamical cloud evolution and the suite of models included in the
current analysis. In Section~\ref{sec:prestellar} we present results
on the angular momentum distribution of starless molecular cloud
  cores. We call them prestellar cores. Some of them collapse to become
protostellar cores. In Section~\ref{sec:protostellar} we investigate
the angular momentum evolution of their central protostellar objects.
We report a statistical correlation between specific angular momentum
and mass, and analyze its dependence on the turbulent environment. The
angular momentum vectors of neighboring protostars tend to be aligned,
at least in the early accretion phase. This is discussed in
Section~\ref{sec:corr}. Finally, in Section~\ref{sec:summary}, we
summarize and conclude.

\section{Models of dynamical cloud evolution}
\label{sec:models}       

\subsection{Numerical method}
\label{subsec:numerics}
To adequately describe the angular momentum evolution during
gravoturbulent fragmentation, it is prerequisite to resolve formation
and subsequent evolution of collapsing fragments over several orders
of magnitude in density. Due to the stochastic nature of supersonic
turbulence, it is not known in advance where and when local collapse
occurs. We therefore resort to {\em smoothed particle hydrodynamics}
(SPH) to solve the equations of hydrodynamics. It is a Lagrangian
method with the fluid represented by an ensemble of particles and flow
quantities obtained by local averaging \citep{BEN90,MON92}. The method
is able to resolve large density contrasts as particles are free to
move and so naturally the particle concentration increases in
high-density regions.  Because it is computationally prohibitive to
treat the cloud as a whole, we concentrate on subregions within the
cloud and adopt periodic boundary conditions \citep{KLE97}. Once the
central region of a collapsing gas clump exceeds a density
contrast of $\sim 10^5$, we introduce a ``sink particle''
\citep*{BAT95}, which accretes gas from its surrounding while keeping
track of mass and linear and angular momentum. Replacing the high-density
cores with ``sink particles'' allows us to follow the angular momentum
evolution of collapsed cores over many free-fall times. 
Altogether, the performance and convergence properties of the
  method are well understood and tested against analytic models and
  other numerical schemes in the context of turbulent
  supersonic astrophysical flows. For a
detailed discussion the reader is referred to \citet{BAT97},
\citet{MAC98}, \citet{LOM99}, \citet{KLE00a,KLE01b} and
\citet{KLE00b}.


\subsection{Models of turbulent self-gravitating clouds}
\label{subsec:turb-models}
The large observed linewidths in molecular clouds imply the presence
of supersonic velocity fields that carry enough energy to
counterbalance gravity on global scales \citep*[see, e.g., the review
by][]{WIL00}.  However, it is known that turbulent energy dissipates
rapidly, roughly on the free-fall timescale \citep{MAC98,STO98,PAD99}.
Unlike previously thought, this is independent of the presence of
magnetic fields. Magnetic fields also do not significantly alter the
efficiency of local collapse for driven turbulence \citep*{HEI01}. To
a first approximation, the fields are therefore taken as being
dynamically unimportant in our current models, and are not included.
We also disregard possible feedback effects from the star formation
process itself (like bipolar outflows, stellar winds, or ionizing
radiation from new-born O or B stars). Our analysis of angular
momentum evolution of protostellar cores therefore focuses on the
process of gravoturbulent fragmentation, i.e.\ on the interplay
between turbulence and self-gravity only.

The suite of models considered here consists of 12 numerical
simulations where turbulence is maintained with constant
root-mean-square Mach numbers in the range $2 \le {\cal M} \le 10$.
This roughly covers the range observed in typical Galactic molecular
clouds.  We apply a non-local scheme that inserts energy in a limited
range of wavenumbers at a given rate \citep{MAC99}. We distinguish
between turbulence that carries its energy mostly on large scales, at
wavenumbers $1 \le k \le 2$, on intermediate scales, i.e.\ $3 \le k
\le 4$, and on small scales with $7 \le k \le 8$. The corresponding
wavelengths are $\ell = L/k$, where $L$ is the total size of the
computed volume.  The models are labeled mnemonically as M$\cal
M$k$k$, with rms Mach number $\cal M$ and wavenumber $k$.  
Each of these
simulations contains 205379 SPH particles. We also
consider a model that is globally unstable and contracts from Gaussian
initial conditions without turbulence \citep[for details
see][]{KLE00a,KLE01b}. It is called GA 
and was run with 500000 particles.  The main parameters are
summarized in Table~\ref{tab:prop}. 
Note, the final star formation efficiency varies between the
  different models, as indicated in Column 5 of Table~\ref{tab:prop}.
  This simply reflects the evolutionary stage at the time when we stop
  the calculation. In some cases the accretion timescale is too long
  to follow the simulation to high efficiencies.
%
%

\begin{table}[h]
\caption{Sample parameters, name of the environment used in the text
  consisting of the Mach number $\mathcal{M}$ and the driving
  scale $k$ (GA denotes the model with Gaussian density), number
  $\mathcal{N}$ of protostellar objects (i.e.\ ``sink particles'' in
  the centers of protostellar cores) at final stage 
  of the simulation, percentage of accreted mass at final stage $
  M_{acc}/M_{tot}$, parameter $\mathcal{A}$ see Eq.~(\ref{eqn:a}), parameter $\mathcal{B}$ see Eq.~(\ref{eqn:b})}
\begin{tabular}{lllllll} \hline \hline
Name & $k$ & $\mathcal{M}$ & $\mathcal{N}$ & $
M_{acc}/M_{tot}$ & $\mathcal{A}$ & $\mathcal{B}$ \\  
& & & & $\left[\%\right]$ &  \multicolumn{2}{c}{$\left[10^{20}\
  \mathrm{cm^2\,s^{-1}}\right]$} \\\hline
M2.0k2 & 1..2 & 2.0 & 68 & 75 & 1.7 & 1.7 \\ 
M2.0k4 & 3..4 & 2.0 & 62 & 48 & 2.0 & 2.0 \\ 
M2.0k8 & 7..8 & 2.0 & 11 & 66 & 1.7 & 1.6 \\ 
M3.2k2 & 1..2 & 3.2 & 62 & 80 & 1.2 & 1.3 \\ 
M3.2k4 & 3..4 & 3.2 & 37 & 82 & 2.0 & 1.9 \\ 
M3.2k8 & 7..8 & 3.2 & 17 & 60 & 2.7 & 2.6 \\ 
M6k2 & 1..2 & 6.0 & 110 & 76 & 1.3 & 1.3 \\ 
M6k4 & 3..4 & 6.0 & 60 & 65 & 1.5 & 1.7 \\ 
M6k8 & 7..8 & 6.0 & 7 & 4 & 1.9 & 1.4 \\ 
M10k2 & 1..2 & 10.0 & 100 & 38 & 1.0 & 1.0 \\ 
M10k4 & 3..4 & 10.0 & 10 & 6 & 2.0 & 1.4 \\ 
M10k8 & 7..8 & 10.0 & 27 & 8 & 1.4 & 1.05 \\ 
GA & ... & ... & 56 & 85 & 1.4 & 1.05 \\ \hline
\end{tabular}
\label{tab:prop}
\end{table}

\subsection{Physical scaling and naming convention}
\label{subsec:scaling}
The models are computed in normalized units using an isothermal
equation of state. Scaled to physical units we adopt a temperature of
11.3$\,$K corresponding to a sound speed $c_{\rm s} = 0.2\,$km\,s$^{-1}$,
and we use a mean density of $n({\rm H}_2) = 10^5\,$cm$^{-3}$, which
is typical for star-forming molecular cloud regions \citep*[e.g.\ in
$\rho$-Ophiuchus, see][]{MOT98}. The mean thermal Jeans mass\footnote{We use a
  spherical definition of the Jeans mass, $M_{\rm J} \equiv 4/3\,\pi
  \rho \lambda_{\rm J}^3$, with density~$\rho$ and Jeans length
  $\lambda_{\rm J}\equiv \left(\frac{\pi{\cal R}T }{G
      \rho}\right)^{1/2}$ and where $G$ and $\cal R$ are the
  gravitational and the gas constant. The mean Jeans mass $\langle
  M_{\rm J} \rangle$ is then determined from the average density in
  the system $\langle \rho \rangle$.  } in all models is $\langle
M_{\rm J} \rangle = 1\,$$M_{\odot}$. The turbulent models contain a
mass of 120$\,$$M_{\odot}$ within a cube of size $0.28\,{\rm pc}$, and
the Gaussian model has 220$\,$$M_{\odot}$ in a volume of
$(0.34\,$pc$)^3$.  The global free-fall timescale is $\tau_{\rm ff} =
10^5\,$yr, and the simulations cover a density range from $n({\rm
  H}_2) \approx 100\,$cm$^{-3}$ in the lowest density regions to
$n({\rm H}_2) \approx 10^9\,$cm$^{-3}$ where the central
parts of   collapsing gas clumps are  converted into 
``sink particles''.

Each ``sink particle'' defines a control volume with a fixed radius of
$560\,$AU. We cannot resolve the subsequent evolution in its interior.
After $\sim10^3\,$yr a protostar will form in the very center.
Because of angular momentum conservation most of the matter that falls
in will assemble in a protostellar disk. It is then transported inward
by viscous and possibly gravitational torques
\citep[e.g.,][]{BOD95,PAP95,LIN96}. With typical disk sizes of order
of several hundred AU, the control volume fully encloses both, star
and disk. If low angular momentum material is accreted, the disk is
stable and most of the material ends up in the central star. In this
case, the disk simply acts as a buffer and smooths eventual accretion
spikes. It will not delay or prevent the mass growth of the central
star by much.  However, if material that falls into the control
volume carries large specific angular momentum, then the mass load
onto the disk is likely to be faster than inward transport. The disk
grows large and may become gravitationally unstable and fragment.
This will lead to the formation of a binary or higher-order multiple
\citep{BOD00}.

Throughout this paper, we adopt the following naming convention: In
the pre-collapse phase, we call high-density gas clumps {\em
  prestellar cores} or simply {\em gas clumps}. They
  build up at the stagnation points of convergent flows. The flows result from
  turbulent motion that establishes a complex network of interacting shocks.
The gas clumps are
identified and characterized using a three-dimensional clump-find
algorithm as described in Appendix A of \citet{KLE00a}. The fluctuations in turbulent velocity fields are highly transient.
They can disperse again once the converging flow fades away.  Even
clumps that are strongly dominated by gravity may get disrupted by the
passage of a new shock front.  For local collapse to result in the
formation of stars, Jeans-unstable, shock-generated density
fluctuations therefore must collapse to sufficiently high densities on
time scales shorter than the typical time interval between two
successive shock passages.  We  include in our
  analysis only Jeans-unstable gas clumps. Angular
momentum is calculated from the internal motions with respect to the
location of the density maximum. These objects correspond to the so
called starless cores observed, e.g., by \citet{GOO93}, \citet{BAR98},
\citet{JIJ99}, and others. They are thought to collapse and build up a
  central protostar or protostellar system in the later stages of evolution. Once collapse has
lead to the formation of an embedded protostar (in our scheme
identified by a central ``sink particle'') we call the object {\em
  protostellar core} or also {\em protostar}.  The angular momentum is
obtained as the spin accumulated by the ``sink particle'' during its
accretion history. The distribution is best compared with observations
of main-sequence binaries as we expect the unresolved star-disk system
interior to the ``sink particle'' to break up into a binary or
higher-order multiple.
      
\section{Molecular cloud clumps and prestellar cores}
\label{sec:prestellar}

\begin{figure}[h]
  \resizebox{\hsize}{!}{\includegraphics{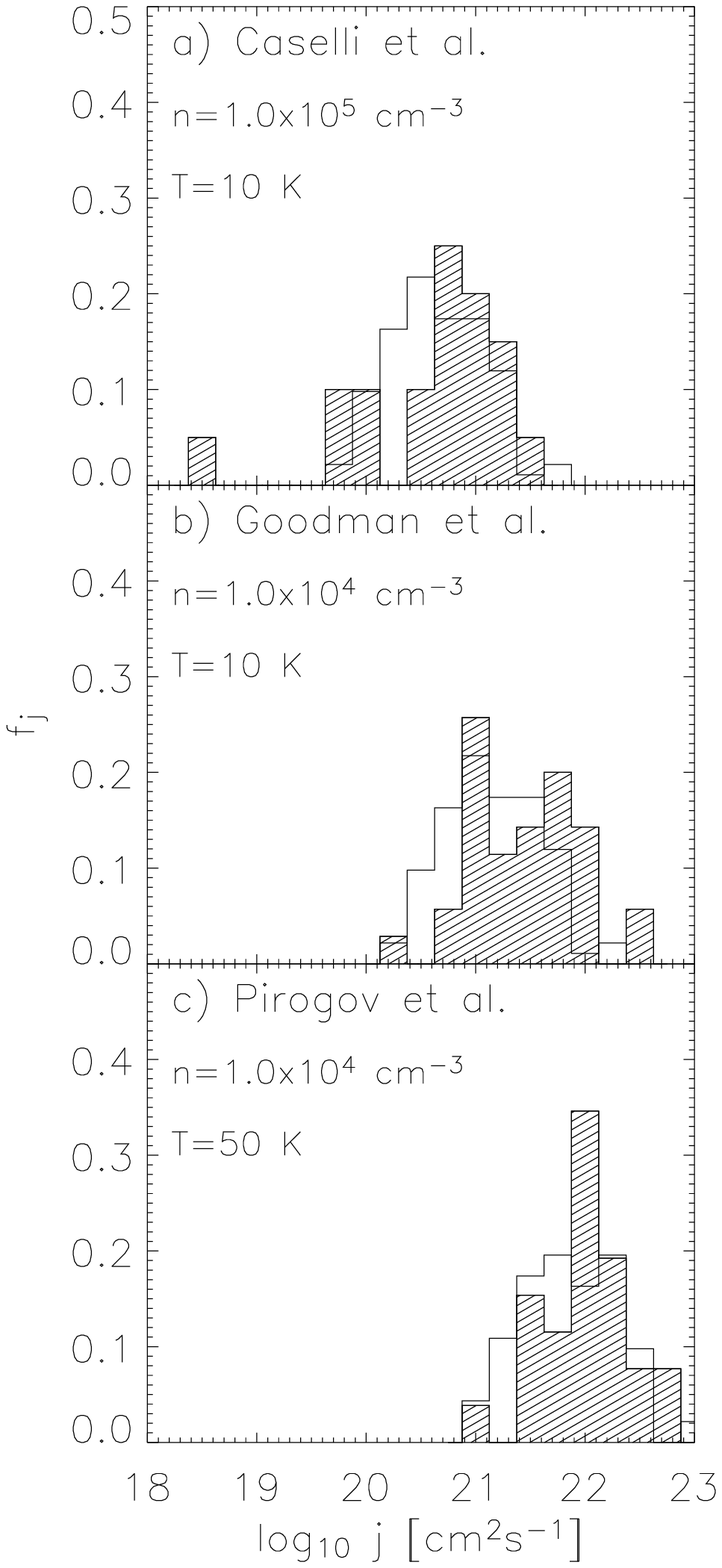}}
\caption{The distribution of specific angular momenta of prestellar cores
  formed in our simulations using model M6k2 (non-hatched histogram)
  is compared to the distribution of specific angular momenta of
  observed molecular cloud cores (hatched distributions). The
  observational data were taken in (a) from Table~5 in \citet{CAS02},
  in (b) from Table~2 in \citet{GOO93}, Table~4 in \citet{BAR98} and
  Table~A2 in \citet{JIJ99} and in (c) from Table~7 in \citet{PIR03}.
  We take $f_{j}$ to represent the percentage of the total number of
  existing cores in a certain specific angular momentum bin.}
\label{fig:clumphisto}
\end{figure}

Figure~\ref{fig:clumphisto} shows the distribution of the specific
angular momenta of the gas-clumps that were identified in the
turbulent environment M6k2 (see Table~\ref{tab:prop}). 
\begin{figure}
  \resizebox{\hsize}{!}{\includegraphics{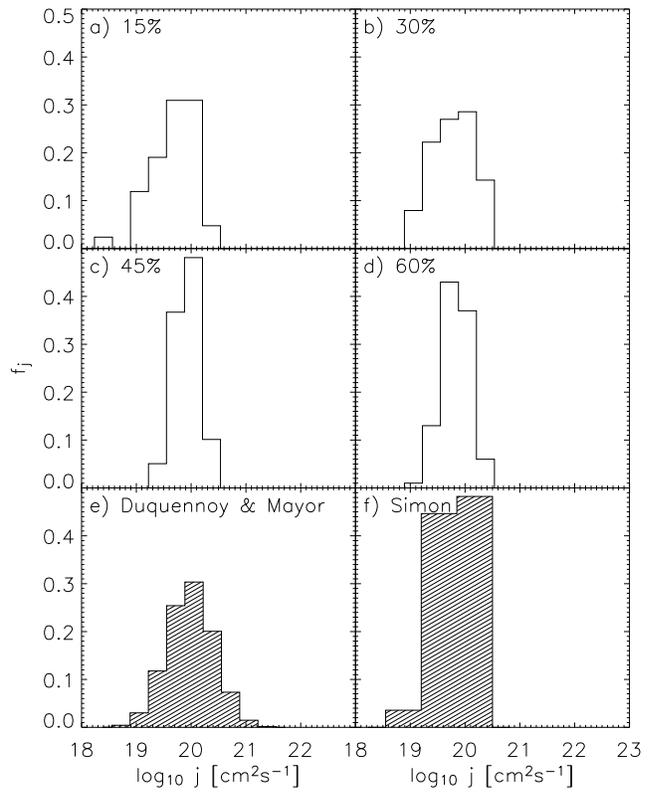}}
\caption{Distribution of specific angular momenta of the protostars or
  protostellar systems at different evolutionary phases of the
  numerical model M6k2 as denoted by the local star formation
  efficiency in percent ((a)-(d)). We compare in (e) with the
  $j$-distribution of binaries among nearby G-dwarf stars from
  \citet{DUQ91b} (for details see text) and in (f) with the
  distribution of specific angular momenta of binaries in the Taurus
  star-forming region from Figure~5 in \citet{SIM92}.  Again, $f_{j}$
  represents the normalized distribution function.}
\label{fig:histo1}
\end{figure}
We compare with observational values taken from \citet{CAS02} for
(a), from \citet{GOO93}, \citet{BAR98}, \citet{JIJ99} for (b) and from \citet{PIR03} for (c). According to \citet{GOO93} the values for the
specific angular momenta are obtained by using best-fit velocity
gradients from maps of observed line-center velocities under the
assumption of solid body rotation. The cores in \citet{GOO93},
\citet{BAR98}, \citet{JIJ99} were mapped in the (J,K)\,=\,(1,1) transition
of $\mathrm{NH_3}$, whereas the massive cloud cores in \citet{PIR03}
and the dense cloud cores in \citet{CAS02} were mapped in
$\mathrm{N_2H^+}$. As found in \citet{CAS02} the two molecular species
trace essentially the same material, especially in starless cores.
Model clumps and observed molecular cloud clumps have comparable mass
spectra \citep{KLE01c} and similar shapes \citep{KLE00a,BAL03}.
Figure~\ref{fig:clumphisto} demonstrates that similar holds for the
distribution of specific angular momenta. 

Note, when transforming from dimensionless code units into physical
scales, the specific angular momentum depends on the mean density~$n$
and the temperature~$T$ as  $j \propto T/\sqrt{n({\rm H}_2)}$. In
Figure~\ref{fig:clumphisto}c, we use our standard scaling
corresponding to regions like the $\rho$-Ophiuchi main cloud \citep{MOT98}. This is
adequate for the low-mass cores studied by \citet{CAS02}, and in
Sect.~\ref{sec:protostellar}, we will furthermore show that the
specific angular momenta of collapsed cores then fall into the right
range for main-sequence binaries. 

We find that the specific angular momentum of prestellar cores has values
between $1 \times 10^{20}\,\mathrm{cm^2\,s^{-1}}$ and $5 \times 10^{21}\, 
\mathrm{cm^2\,s^{-1}}$ with a mean value of approximately $\langle j
\rangle\approx 5 \times 10^{20}\,\mathrm{cm^2\,s^{-1}}$.
This is in good agreement with
the \citet{CAS02} sample which has $\langle j \rangle =7\times10^{20}\,\mathrm{cm^2\,s^{-1}}$. Their cloud cores have a mean mass of $\sim
6\,M_{\odot}$ comparable to the core masses in our
simulations.  A Kolmogorov-Smirnov (KS) test was performed. We find
that at a $50\,\%$ level the distributions are statistically
indistinguishable.

However, the cores observed by \citet{GOO93} and \citet{PIR03} trace
lower densities and have higher mean masses of around $50\, M_{\odot}$
and $500\, M_{\odot}$, respectively.  In Figure~\ref{fig:clumphisto}b
we therefore use $n({\rm H}_2) = 1 \times 10^4 \,{\rm cm}^{-3}$ and
$T=10\,$K leading to $\langle j \rangle =
1.5\times10^{21}\,\mathrm{cm^2\,s^{-1}}$. This matches the the
observations since {\citet{GOO93}} find $\langle j \rangle =
6\times10^{21}\,\mathrm{cm^2\,s^{-1}}$ and a median value of $3\times10^{21}\,\mathrm{cm^2\,s^{-1}}$. The massive cores mapped by
\citet{PIR03} have higher velocity dispersions, higher kinetic
temperatures ($20-50~\mathrm{K}$) and a density~$n({\rm H}_2)\ge
1 \times 10^4 \,{\rm cm}^{-3}$. The resulting mean specific angular
momentum is $\langle j \rangle =
1.5\times10^{22}\,\mathrm{cm^2\,s^{-1}}$.  Again, adequate scaling in
Figure~\ref{fig:clumphisto}a results in higher values of $j$ in our
models and leads to better agreement with the observed distribution.

 Given the simplified assumptions in our numerical models, we
  altogether find remarkably good agreement with the observed specific
  angular momenta in the prestellar phase. Similar findings are
  reported by \citet{GAM03}. Similar to our study, they follow the
  dynamical evolution of isothermal, self-gravitating, compressible,
  turbulent ideal gas. However, they include the effects of magnetic
  fields and solve the equations of motion using a grid-based method
  (the well-known ZEUS code). Their approach is thus quite complementary to
  ours.  The $j$ distribution that results from their simulations
  peaks at $4\times10^{22}\,\mathrm{cm^2\,s^{-1}}$. They fix the mean
  number density at $n({\rm H}_2)\approx 1.0 \times 10^2 \,{\rm
  cm}^{-3}$ and use  $T=10~\mathrm{K}$. Using the same physical scaling
  we get very similar values, i.e.\ $\langle j \rangle =
  4\times10^{22}\,\mathrm{cm^2\,s^{-1}}$.

  This mean value also falls into the range of specific angular momenta of cores that form in simulations by \citet{LI04}. 
  They also use a version of the ZEUS code (ZEUS-MP) to perform
  high-resolution, three-dimensional, super-Alfv{\' e}nic turbulent
  simulations in order to investigate the role of magnetic fields in
  self-gravitating core formation within turbulent molecular clouds. Adopting the
  same physical scaling as in \citet{GAM03}, the specific angular momentum of
  their cores takes values between $5\times10^{21}\,\mathrm{cm^2\,s^{-1}}$ and
  $8\times10^{22}\,\mathrm{cm^2\,s^{-1}}$.

\section{Protostars and protostellar systems}
\label{sec:protostellar}
Figure~\ref{fig:histo1} shows the distribution of specific angular
momentum of collapsed cores at four different stages of mass
accretion, ranging from $15\,\%$ of the total available mass in the
molecular cloud accreted onto collapsed cores in
Figure~\ref{fig:histo1}a to $60\,\%$ in Figure~\ref{fig:histo1}d.  While
the distribution narrows during the evolution, the mean specific
angular momentum remains essentially at the same value
$j=(8\pm2)\times 10^{19}\,\mathrm{cm^2\,s^{-1}}$ with a range from
$10^{18}\, \mathrm{cm^2\,s^{-1}}$ to $5 \times 10^{20}\, \mathrm{cm^2\,s^{-1}}$.
The specific angular momentum of the protostellar cores in the
considered model are approximately one order of magnitude smaller than
the ones of the Jeans-unstable clumps, but both distributions join
without gap.  In a statistical sense, there is a continuous transition
as loss of angular momentum occurs during contraction.  The range of
specific angular momenta of the protostellar cores agree 
well with the observed values for binaries \citep[e.g.,][]{BOD95}.
For this reason we compare in Figures~\ref{fig:histo1}a--d the model
distributions with observations of binaries among G-dwarf stars by
\citet{DUQ91b} in Figure~\ref{fig:histo1}e and with observations of
young star binaries in the Taurus star forming region by \citet{SIM92}
in Figure~\ref{fig:histo1}f.

\citet{DUQ91b} derived a Gaussian-type period distribution for their
sample. Based on this distribution we calculated the distribution of
the specific angular momenta using the following equation \citep[see
also][ Eq.~10]{KRO95b}:
\begin{displaymath}
j=6.23 \times 10^{18} \left(1-e^2\right)^{1/2} P^{1/3}
\frac{m_1 m_2}{\left(m_1 + m_2 \right)^{4/3}}\left(\mathrm{cm^2\,s^{-1}}\right)
\label{eqn:jper}
\end{displaymath}
where masses are in $M_{\odot}$ and $P$ in days.  We used a
primary mass $m_1$ of $1\,M_{\odot}$, a mean mass ratio
between primary and secondary $q=m_2/m_1=0.25$ and a mean eccentricity
$e=0.31$.

The resulting distribution has a mean specific angular momentum of
$1.6\times10^{20}\, \mathrm{cm^2\,s^{-1}}$, which agrees well with the
values from the simulations. This also holds for
Figure~\ref{fig:histo1}f which was taken from Figure~5 in
\citet{SIM92}. The mean specific angular momentum here has a value of
$1.6\times10^{20}\, \mathrm{cm^2\,s^{-1}}$.  The agreement of the
distributions was confirmed by a ${\chi}^2$ statistical test.  Since
our numerical resolution is not sufficient to follow the
subfragmentation of collapsing cores into binary or higher-order
multiple systems, the time evolution of $j$ is an important tool to
evaluate or our models.  We see a clear progression from the
rotational properties of gas clumps (as discussed in
Sect.~\ref{sec:prestellar}) to those of the resulting collapsed cores.
A similar correlation is observed between the cloud core \citep{CAS02}
and typical main-sequence binaries \citep{DUQ91b}.  The former may be
the direct progenitors of the latter.

\begin{figure*}
\sidecaption
\resizebox{\hsize}{!}{\includegraphics{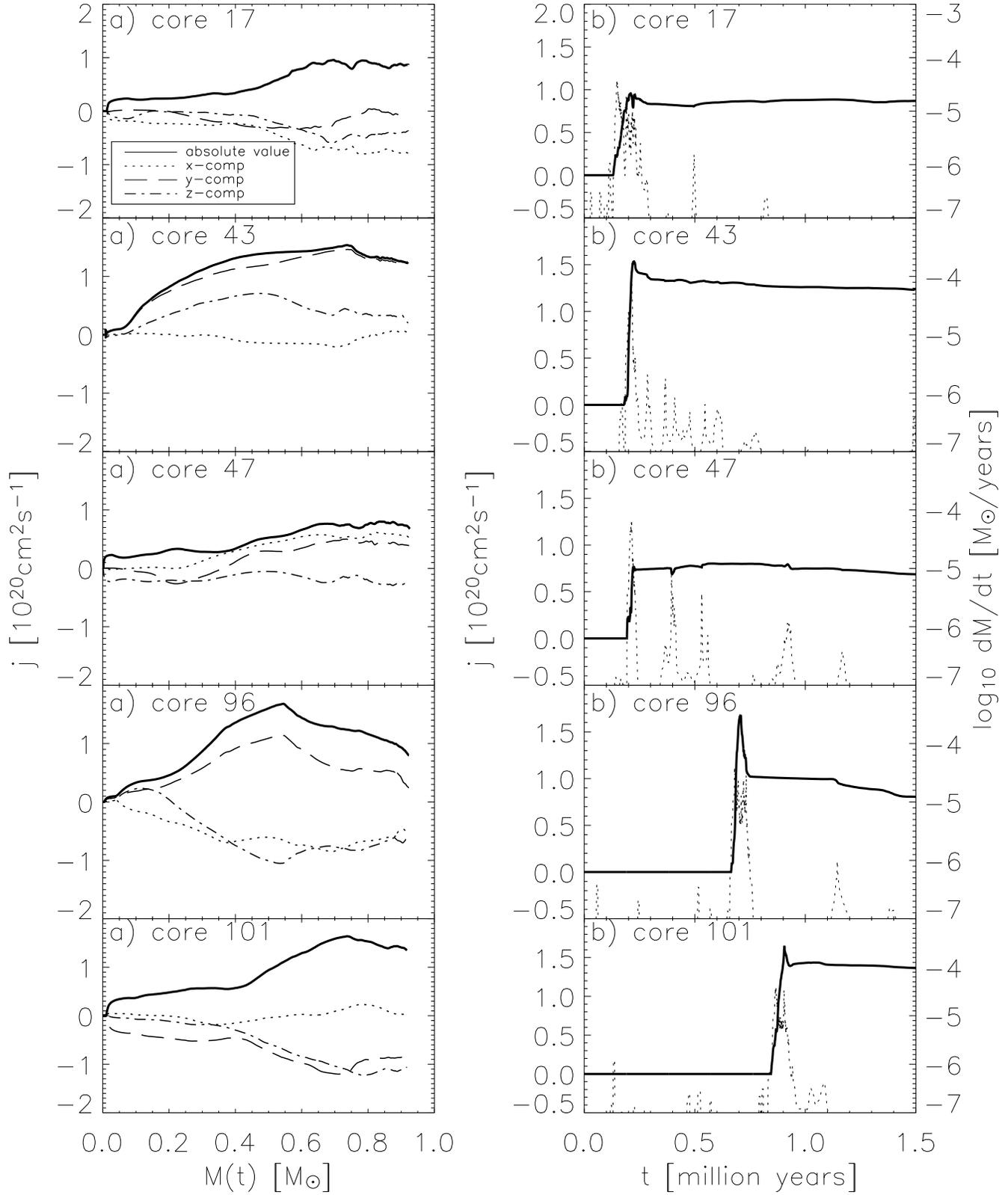}}
\caption{Absolute value of specific angular momentum (\textit{solid
    line\nocorr}) of our model M6k2 as (a) a function of mass and as
  (b) a function of time for five different protostellar objects (1-5) with
  approximately equal final masses ($M=0.94\,M_{\odot}$). In (a)
  the x-component (\textit{dotted line\nocorr}), the y-component
  (\textit{dashed line\nocorr}) and the z-component
  (\textit{dashed-dotted line\nocorr}) of the specific angular momenta
  are shown as well. For comparison the mass accretion rates onto
  the protostar are indicated in (b) by a dashed line (\textit{associated
    y-axis on the right hand side\nocorr}).}
\label{fig:jt1}
\end{figure*}

\subsection{Example of the angular momentum  evolution of a protostellar
  core in a cluster}

The evolution of the specific angular momentum of individual
protostellar cores can be very complex depending on the rotational
properties of their environment. There is a strong connection to the
time evolution of the mass accretion rate.

In Figure~\ref{fig:jt1} we select five collapsed cores in model M6k2
with about the same final mass.  All of them show a similar evolution
the specific angular momentum with increasing mass
(Fig.~\ref{fig:jt1}a) and time (Fig.~\ref{fig:jt1}b). Nevertheless
there are visible differences in the details.

Initially, the specific angular momentum increases with growing mass.
However, at later stages the evolution strongly depends on the secular
properties of the surrounding flow.  In cores~43, 96, and 101, for
example, $j$ decreases again after reaching a peak value, while for
cores~17 and 47 $j$ stays close to the maximum value.  Depending on
the specific angular momentum of the accreted material the resulting
protostellar disks are expected to evolve quite differently.  For
example, preliminary 2-dimensional hydrodynamic calculations, show
that core~17 and core~47 will probably develop a stable disk
(Bodenheimer 2003, priv. comm.). The ratio of rotational to
gravitational energy for the peak value of $j$ is $\beta=0.005$ for
core~17 and $\beta=0.003$ for core~43.  On the other hand, core~101
will fragment into a binary star. It has a $\beta=0.016$. Also, the
disk of core~96 is highly unstable with corresponding $\beta=0.016$.
The evolution of core~43 has not yet been followed sufficiently long
to determine whether it will fragment to form a binary star or not.
These results show the importance of the specific angular momentum on
the evolution of the protostellar object. A high value $\beta > 0.01$
leads to fragmentation whereas $\beta < 0.01$ results in a stable
disk.  A similar result was found by \citet{BOS99} for slowly
rotating, magnetic clouds.

Figure~\ref{fig:jt1}b shows, that the change in specific angular
momentum is closely linked to the mass accretion. At the point in time
where the mass accretion rate (dotted line) has a pronounced peak, the
change in specific angular momentum is also significant. A high mass
accretion rate can result in an increase of specific angular momentum.
But as seen for core~96 in Figure~\ref{fig:jt1}b a high mass accretion
rate can also lead to a reduction of specific angular momentum. 
The exact evolution of the specific angular momentum is thus closely
linked to the flow properties of the surrounding material.

\subsection{Statistical correlation between specific angular
  momentum and mass}

\label{sec:massj}
\begin{figure}
  \resizebox{\hsize}{!}{\includegraphics{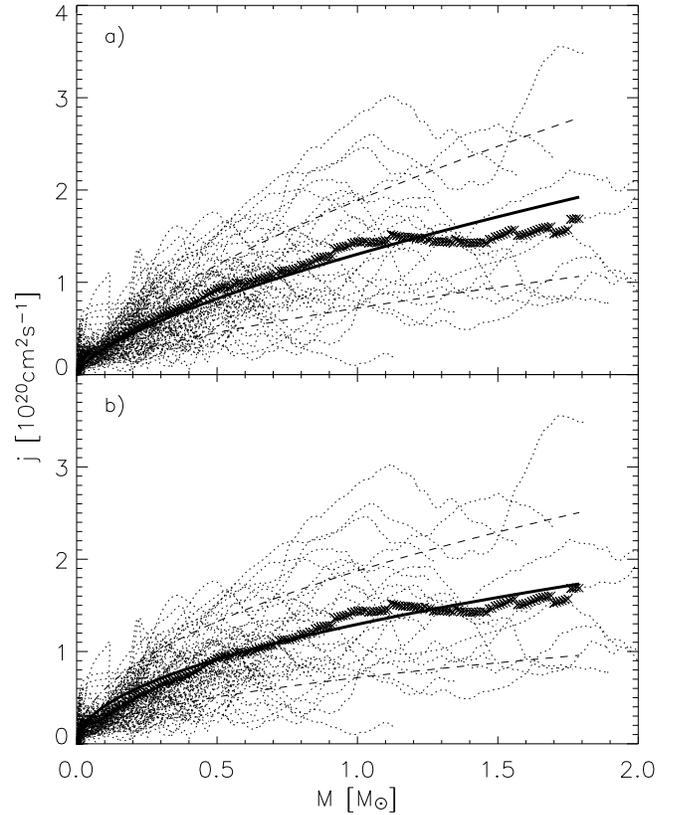}}
\caption{Absolute values of specific angular momenta (\textit{dotted
    lines\nocorr}) of all protostars (i.e.\ ``sink particles'') from
  our model M6k2 as a function of mass. The specific angular momenta
   are averaged at certain mass values which are separated
  by $0.01\,M_{\odot}$ and the resulting points are indicated
  by crosses. The \textit{solid line} represents a fit of these
  averaged specific angular momenta in the mass range between $0$ and
  $1.7\,M_{\odot}$ In (a) we fit with a function of the
  form: $j=\mathcal{A} {(M/ M_{\odot})^{2/3}}$,
  $\mathcal{A}=(1.3\pm0.6)\times10^{20}\,\mathrm{cm^2\,s^{-1}}$, and in
  (b) we use a square root function: $j=\mathcal{B} \sqrt{M/
    M_{\odot}}$, $\mathcal{B}=(1.3\pm0.6)\times10^{20}\,
  \mathrm{cm^2\,s^{-1}}$. One standard deviation is marked by the
  \textit{dashed lines}.}
\label{fig:alljm1}
\end{figure}

\begin{figure*}
\sidecaption
\resizebox{\hsize}{!}{\includegraphics{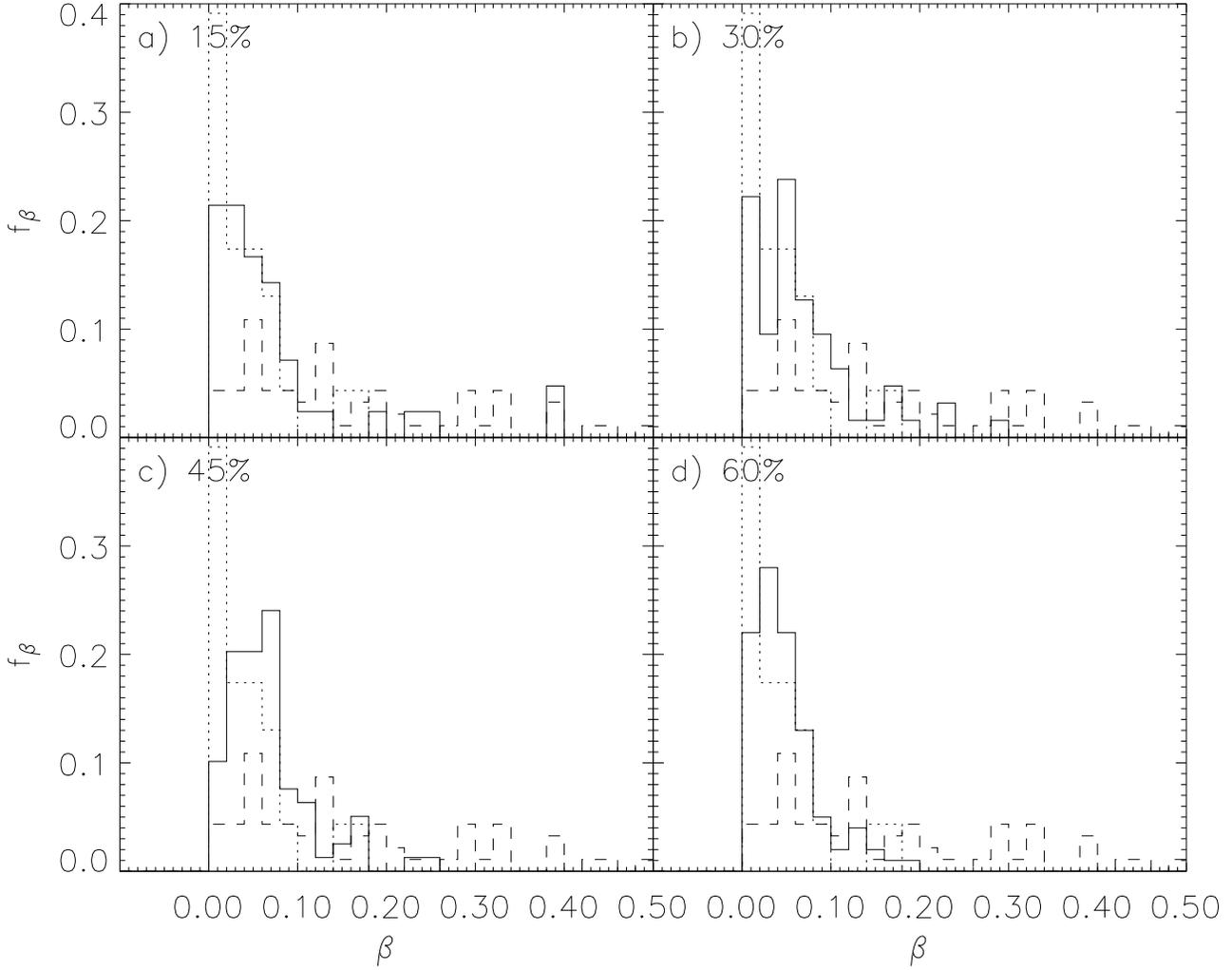}}
\caption{Distribution of $\beta$ obtained from our model M6k2 with $15\,\%$ (a),
  $30\,\%$ (b), $45\,\%$ (c) and $60\,\%$ (d) of the material accreted.
  For the collapsed protostellar cores in our models
  (\textit{solid lines}) we calculate $\beta$  
   using Eq.~\ref{eqn:energy}, where we  assume  solid-body rotation of
  a uniform density sphere and take the calculated specific angular
  momentum $j$. For the prestellar cores (i.e.\ Jeans-unstable gas
  clumps; see the \textit{dashed lines})
  we derive  $\beta$ self-consistently from the three-dimensional density
  and velocity structure (using Eq.~\ref{eqn:beta3}). For comparison
  we also indicate with \textit{dotted lines} the values reported by \citet[][ see their
  Figure~11]{GOO93} for observed  molecular cloud cores. These were
  obtained with the same assumptions,  
  and we use the same binning with   $f_{\beta}$ representing the fraction
  of objects per
  $\beta$ bin.}
\label{fig:beta}
\end{figure*} 
As mentioned above we find a correlation between mass~$M$ and specific
angular momentum~$j$ in a {\em statistical sense}.  The result is depicted
in Figure~\ref{fig:alljm1}.  It shows the angular momentum evolution
as function of mass for all 110 collapsed cores in model M6k2. 

In
Figure~\ref{fig:alljm1}a, following \citet{GOO93}, we adopt rigid body rotation with constant angular velocity~$\Omega$ and and uniform core density~$\rho$.
With these assumptions the specific angular momentum~$j$ can be written as:
\begin{equation}
j=p\Omega R^2
\label{eqn:a1}
\end{equation}
For a uniform density sphere $p=\frac{2}{5}$.  The mass~$M$ of a
sphere with constant density~${\rho}_0$ is related to the radius~R
via:
\begin{equation}
M=\frac{4\pi}{3}{\rho}_0 R^3
\label{eqn:a2}
\end{equation}
From Eq.~\ref{eqn:a1} and Eq.~\ref{eqn:a2} follows that $j$ can be
expressed as:
\begin{equation}
j=p\Omega\left(\frac{3}{4\pi{\rho}_0}\right)^{2/3}M^{2/3}
\label{eqn:a3}
\end{equation}
Therefore we fit the average angular momentum with a function of the form:
\begin{equation}
j=\mathcal{A}{(M/ M_{\odot})^{2/3}},
\label{eqn:a}
\end{equation} 
where ${\mathcal
  A}=p\Omega\left(\frac{3{M}_{\odot}}{4\pi\rho_0}\right)^{2/3}$.
From Figure~\ref{fig:alljm1}a the constant ${\mathcal
  A}$ has a value of $(1.3\pm0.6)\times10^{20}\, \mathrm{cm^2\,s^{-1}}$. This fit formula can be applied to
different turbulent cloud environments, and we list the corresponding
values of $\mathcal{A}$ for our model suite in Table~\ref{tab:prop}.
Using the fitted value and the
density~${{\rho}_0}=4\times{10}^{-15}\,\rm{g\,cm^{-3}}$ where protostellar cores are
identified, we calculate an angular
velocity~$\Omega=1.33\times{10}^{-11}\,\rm{s^{-1}}$.

In this picture the ratio of
rotational to gravitational energy~$\beta$ can be written as:
\begin{equation}
 \beta  = \frac{(1/2)I\Omega^2}{qGM^2/R}=\frac{3p}{8 \pi q}\frac{{\Omega}^2}{{\rho}_0G},
\label{eqn:energy}
\end{equation}
where the moment of inertia is given by $I = pMR^2$, and
$q=\frac{3}{5}$ is defined such that $qGM^2/R$ represents the
gravitational potential of a uniform density sphere.
With the assumptions of constant angular velocity and uniform density it
follows that $\beta$ is also constant. With $\Omega$ and $\rho_0$ as above, we
get values $\beta\approx0.05$. 

\citet{GOO93} as well as \citet{BUR00} derived scaling relations where
$\beta$ is independent of radius. Similar values for $\beta$ were also found by \citet{GOO93} for the
observed cloud cores. In good agreement with our calculations they
found that all values are below $0.18$ with the majority under $0.05$.

The fit in Figure~\ref{fig:alljm1}a rests on the assumption of the
collapse of an initially homogeneous sphere with constant angular
velocity.  Using ``sink particles'' however, which have a constant
radius, makes it necessary to examine another possibility.  In
Figure~\ref{fig:alljm1}b we thus assume a constant radius~$R$ and a
constant $\beta$. Choosing a constant $\beta$ is supported by the
observations as discussed above and by our simulations as we show below.  With $\beta$ and
Eq.~\ref{eqn:energy} (which still holds) it follows  that the angular
velocity depends on the density as
\begin{equation}
\Omega=\sqrt{\frac{8 \pi q}{3p}G\rho\beta}\;.
\label{eqn:b1}
\end{equation} 
Thus, the angular velocity~$\Omega$ is no longer a constant. This implies for the specific angular momentum~$j$:
\begin{equation}
j=p\Omega R^2=\sqrt{2pRqG\beta}\sqrt{M}
\label{eqn:b2}
\end{equation}
Following Eq.~\ref{eqn:b2} we fit our data in Figure~\ref{fig:alljm1}b
 with a square root function:
\begin{equation}
j=\mathcal{B}\sqrt{M/ M_{\odot}}\;,
\label{eqn:b}
\end{equation} 
where the moment of inertia is given by $I = pMR^2$, and
$q=\frac{3}{5}$ is defined such that $qGM^2/R$ represents the
gravitational potential of a uniform density sphere. With the above
density and angular velocity Eq.~\ref{eqn:energy} results in
$\beta\approx0.05$. For the
new fit we find a scaling factor $\mathcal{B} =
(1.3\pm0.6)\times10^{20}\, \mathrm{cm^2\,s^{-1}}$ in the mass range $0 \le
M/{M}_{\odot} \le 1.7$. Again, this fit formula can be applied to
different turbulent cloud environments, and we list the corresponding
values of $\mathcal{B}$ for our model suite in Table~\ref{tab:prop}.

The question remains if our simulations support the assumption of a constant
$\beta$. In Figure~\ref{fig:beta} we compare the distribution of 
$\beta$ measured  by \citet{GOO93} with values we extract from our
model. For the prestellar cores, we use the definition
\begin{equation}
\beta = \frac{E_{\mathrm{rot}}}{E_{\mathrm{grav}}}\;,
\label{eqn:beta3}
\end{equation}
and calculate rotational and potential energy, $E_{\mathrm{rot}}$ and
$E_{\mathrm{grav}}$, consistently from the full three-dimensional gas
distribution of each gas clump. We do not adopt any assumption about
symmetry and shape of the density and velocity structure.
If we assume each clump is spherical and has roughly constant density,
as implied for example by Eq.~\ref{eqn:energy}, then $\beta$ is
overestimated by a factor of 2.7.  This shows the importance of taking
the full three-dimensional clump structure into account when analyzing
the rotational properties of molecular cloud cores.  Both prestellar
cores (i.e\ Jeans-unstable gas clumps) as well as protostellar cores
in our model typically have $\beta < 0.3$ with similar
distributions. Thus, $\beta$ stays mainly in the interval $[0,0.3]$ and in this sense it remains approximately constant during the collapse.

It should be noted in passing, that we also looked at the density
structure of purely hydrodynamic turbulence, i.e.\ without
self-gravity. If we again perform a clump-decomposition of the density
structure and compute hypothetical $\beta$-values, we find $\beta
\approx 2$. This is indicative of the high degree of vorticity inherent to
all turbulent flows. However,  it also suggests that dense 
molecular cloud cores
are strongly influenced by self-gravity. The fact that all 
cores in the observational sample have  $\beta < 0.2$   implies that
gravitational contraction is needed to achieve density contrasts high
enough for sufficiently low $\beta$ . This agrees with the
picture of gravoturbulent fragmentation where molecular cloud
structure as whole is dominated by supersonic turbulence but stars can
only form in those regions where gravity overwhelms all other forms of
support.
 
\begin{figure*}
\sidecaption
\resizebox{\hsize}{!}{\includegraphics{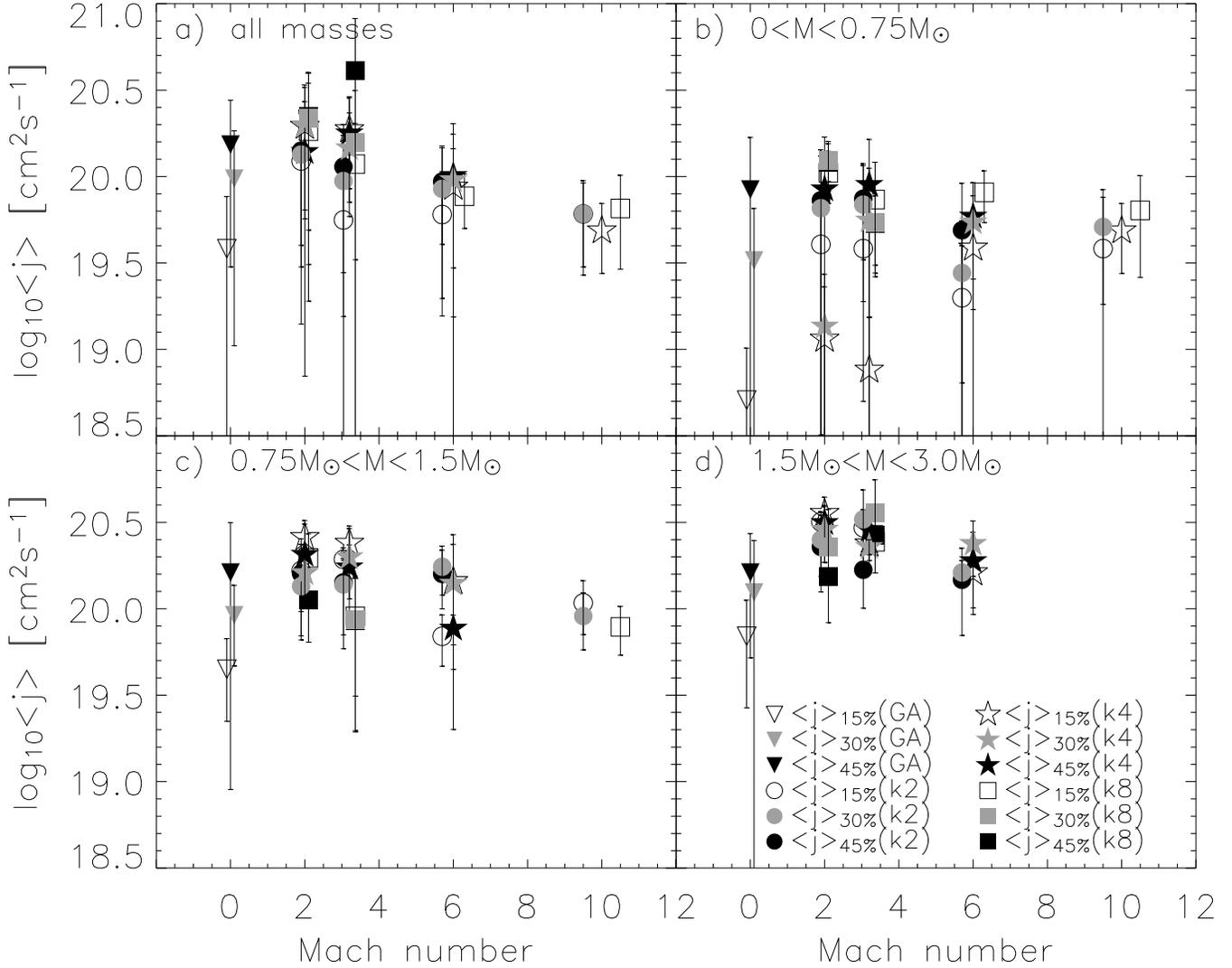}}
\caption{ Average specific angular momentum of
  protostellar objects in different turbulent environments as a
  function of the associated Mach number. Different shapes mark
  different driving scales $k$ (\textit{circle} - $k=2.0$,
  \textit{star} - $k=4.0$, \textit{square} - $k=8.0$). GA stands for
  the Gaussian collapse without driving (\textit{arrow pointing
    downward}). Different colors represent different stages of
  accretion (\textit{white} - 15\,\% material accreted, \textit{gray} -
  30\,\% material accreted, \textit{black} - 45\,\% material accreted).
  In a) all protostars (identified as ``sink particle'' in the
  simulations) were used in calculating the average, in b) through d)
  only objects in the denoted mass bins were considered. The error bars
  show the standard deviation of $<j>$. For more clarity the symbols
  are distributed around the corresponding Mach number $\mathcal{M}$
  ($\frac{\Delta \mathcal{M}}{\mathcal{M}}=5\,\%$)}
\label{fig:mach}
\end{figure*}  

Comparing the two fits in Figure~\ref{fig:alljm1} shows that our first set of
assumptions is a better representation of the data. This is especially
true during the early accretion phase where we have good statistics. And it
applies to different turbulent cloud environments, as well.  We conclude that
-- in a statistical sense -- the angular momentum evolution of collapsing
cloud cores can be approximately described as contraction of initially
constant-density spheres undergoing rigid body rotation with constant angular
velocity. This is consistent with the fact that cores from gravoturbulent
fragmentation follow a Bonnor-Ebert-type radial density profile \citep{BAL03}
and have roughly constant density in their innermost regions,
it also supports the assumptions adopted by
\citet{GOO93} and \citet{BUR00}.

\subsection{Dependence of the specific angular momentum on the environment}

When we compare the results of our complete suite of numerical models
(see Table~\ref{tab:prop}) we find as a general trend that the average
angular momentum falls with increasing Mach number. This is
illustrated in Figure~\ref{fig:mach}a.  However, this follows mainly
from a positive dependence of angular momentum on mass and from 
a correlation between average mass of the cloud cores and Mach number.
As seen in Sect.~\ref{sec:massj} the specific angular momentum
increases on average as the mass of the core rises.  In environments
with a low Mach number the mass growth of the cores is undisturbed
over longer periods of time and so larger masses can accumulate. This
can be inferred from   Table~\ref{tab:prop}, where we list both  the number of
cores and the accreted mass. For higher Mach numbers more
cores with on average less mass form.  Thus, the average angular
momentum is expected to decrease with increasing Mach number
(Fig.~\ref{fig:mach}a).

To detect a direct dependence of the specific angular momentum on the
Mach number we select cores that belong in a certain mass bin and
average the specific angular momentum only over those cores. The
results are shown in Figures~\ref{fig:mach}b-\ref{fig:mach}d.  We find
that there is in general little spread of specific angular momentum
for different scales and different times in accretion history
independently of the Mach number. Low mass protostars
(Fig.~\ref{fig:mach}b) are an exception, in low Mach number
environments they show especially low angular momentum in the early
accretion phase. However, within the error bars we do not find a
further dependence of $j$ on the Mach number.

Compared to the Gaussian collapse case, turbulent driving with small
Mach numbers results in higher specific angular momenta. This is due
to input of turbulent energy that can be converted into rotational
energy if the turbulent velocities are not too high.
 \begin{figure*}
\sidecaption
\resizebox{\hsize}{!}{\includegraphics{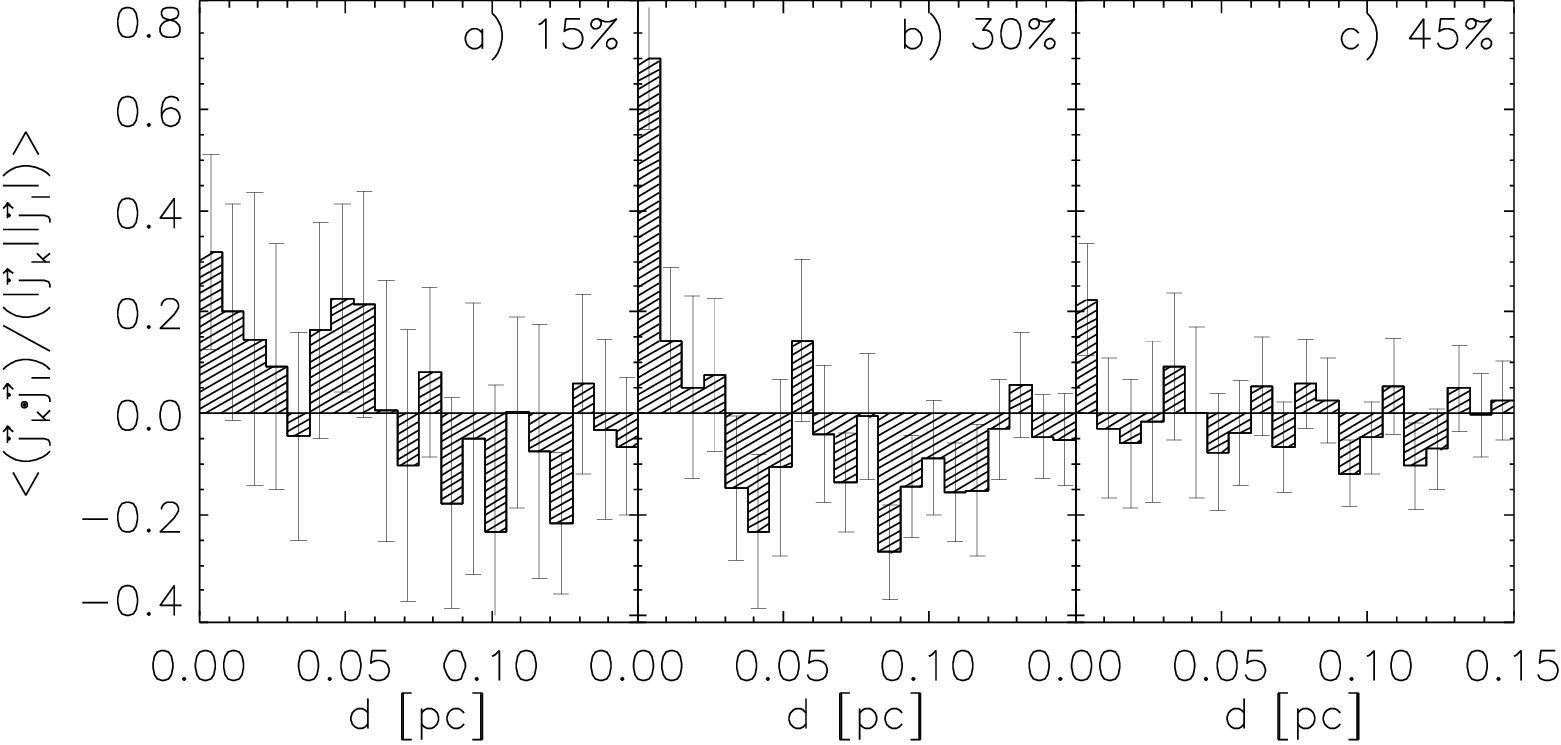}}
\caption{The correlation of specific angular momenta of different protostellar
  cores in model M6k2 with respect to their orientations as a function
  of distance between the cores. As a measure for the correlation the
  scalar product of different cores was taken and averaged over cores
  that exhibit similar distances between each other. High positive
  values denote co-aligned and high negative values denote
  anti-aligned angular momenta. The three graphs (a)-(c) show three
  different times at which $15\,\%$ (1), $30\,\%$ (2) and $45\,\%$ (3) of
  the available material was accreted. The error bars show the
  standard deviation of the averaged correlations. 
  }
\label{fig:corr}
\end{figure*}  
\begin{figure*}
\centering
\resizebox{\hsize}{!}{\includegraphics{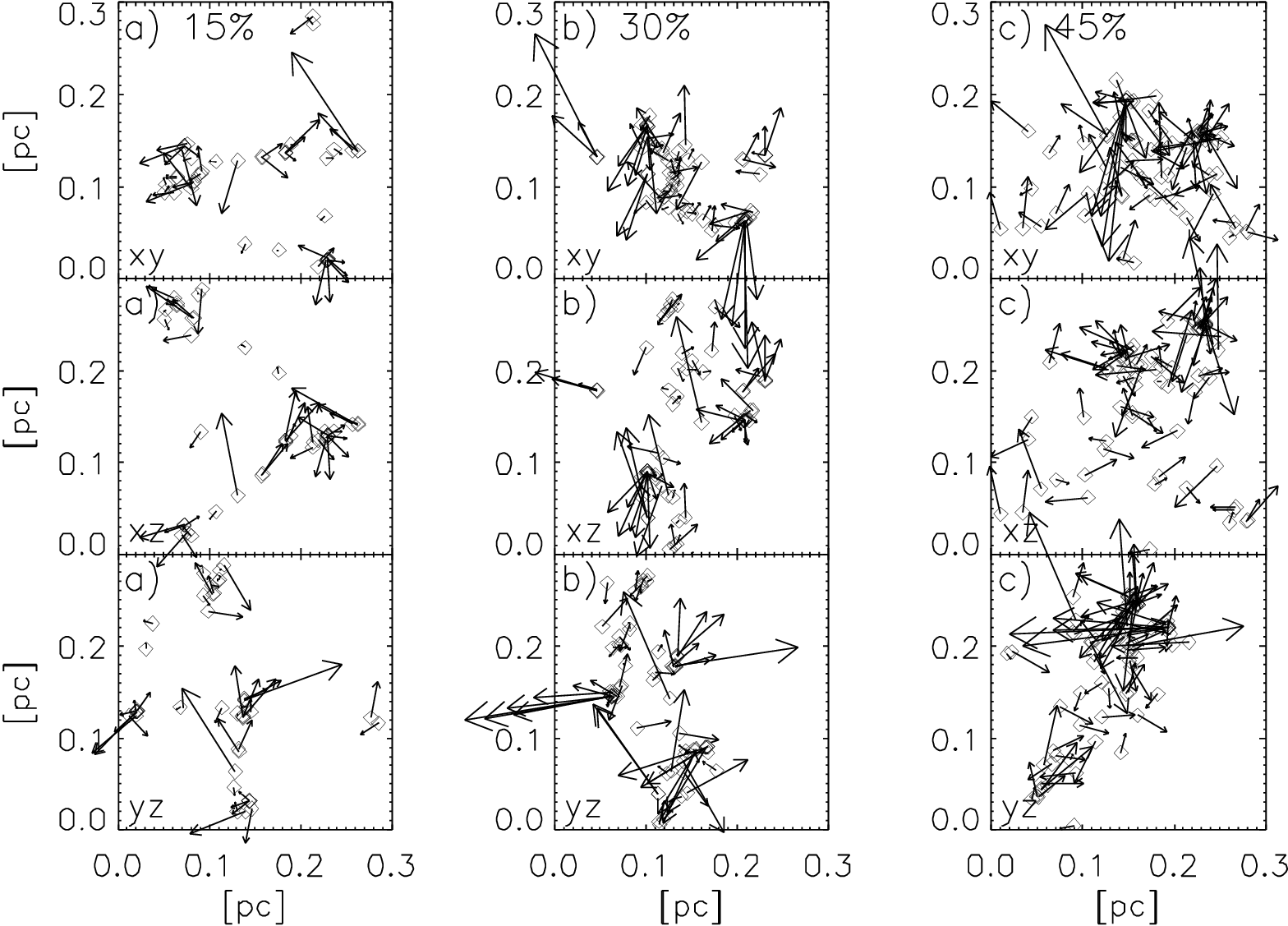}}
\caption{Orientation of the angular momenta (\textit{arrows}) and spatial distribution of the
  protostellar cores (\textit{diamonds}) that formed in model M6k2. We
  present the projections on the xy, xz and yz-plane. The spatial
  distributions are compared for three different times at which $15\,\%$
  (a), $30\,\%$ (b) and $45\,\%$ (c) of the available material are
  accreted. The length of the arrows scales with the specific angular
  momentum of the protostellar core.}
\label{fig:3d}
\end{figure*}         

\section{Orientation of angular momentum vector}
\label{sec:corr}
We find in our simulations that neighboring protostellar cores have
similarly oriented angular momenta.  In Figure~\ref{fig:corr} the
correlation of specific angular momenta of different protostellar
cores with respect to their orientations is shown as a function of
distance. We calculate the scalar product $\vec{j}\cdot \vec{j}$ and
average over cores with similar distances.

Figure~\ref{fig:3d}a shows the spatial configuration in model M6k2
after $15\,\%$ of the available material has been accreted onto the
protostellar cores. These cores form in small aggregates with
diameters below 0.07~pc. 
The corresponding Figure~\ref{fig:corr}a shows
a spatial correlation of the specific angular momenta for small
distances. The correlation length is approximately 0.05~pc.  Thus,
correlation length and cloud size are closely connected. This can be
understood because within one molecular cloud clump neighboring cores
accrete from the same reservoir of gas and consequently gain similar
specific angular momentum.  In the early phase of accretion we
therefore expect  disks and protostellar outflows of
neighboring protostars to be closely aligned.

Indeed, several examples of parallel disks and outflows have been
reported in low-mass, isolated Bok globules by \citet{FRO03},
\citet{KAM03}, \citet{NIS01}, and \citet{SAI95}.  Alternative
explanations for the alignment of the symmetry axes of young stars
include density gradients in the prestellar phase or the presence of
strong magnetic fields. However, \citet{DUC03} found that the disks of
T-Tauri stars driving jets or outflows are perpendicular to the
magnetic field but disks of T-Tauri stars without jet are parallel to
the field lines. This is very puzzling, showing the complexity of the
situation that will naturally arise in strongly turbulent flows.

During subsequent accretion the correlation length decreases to values
below 0.015~pc (Fig.~\ref{fig:corr}b).  This means that only close
systems maintain correlated 
(see also Fig.~\ref{fig:3d}b). This has
three reasons. First, small $N$ systems of embedded cores are likely
to dissolve quickly as close encounters lead to ejection
\citep[e.g.,][]{REI01}. Only close binaries are able to survive for a
long time \citep[e.g.,][]{KRO95a,KRO95b}. The correlation length
therefore decreases with time. Second, the same turbulent flow that
generated a collapsing high-density clump in the first place may also
disrupt it again before it is fully accreted. If the clump contains
several protostars they will disperse, again decreasing the
correlation. Third, the opposite may happen. Turbulence may bring in
fresh gas. The protostars are then able to continue accretion, but the
specific angular momentum of the new matter is likely to be quite
different from the original material.  As protostars accrete at
different rate, we expect a spread in $\vec{j}$ to build up, the
alignment will disappear.  Altogether, at later stages of the
evolution, we expect the correlation between the specific angular
momenta of close protostellar objects is disappeared almost
completely. This is evident in Figure~\ref{fig:corr}c. Furthermore,
Figure~\ref{fig:3d}c demonstrates that most of the initial
subclustering has disappeared by then.

\section{Summary and conclusions}
\label{sec:summary} 

We studied the rotational properties and time evolution of the
specific angular momentum of prestellar and protostellar cores formed
from gravoturbulent fragmentation in numerical models of
supersonically turbulent, self-gravitating molecular clouds. We
considered rms Mach numbers ranging from 2 to 10, and turbulence that
is driven on small, intermediate, and large scales, as well as one
model of collapse from Gaussian density fluctuations without any
turbulence. Our sample thus covers a wide range of properties observed
in Galactic star-forming regions, however, our main focus lies in
typical low- to intermediate-mass star-forming regions like
$\rho$-Ophiuchus or Taurus.

With the appropriate physical scaling, we find the specific angular
momentum~$j$ of prestellar cores in our models, i.e.\ cloud cores as
yet without central protostar, to be on average $\langle j \rangle =
7\times10^{20}\,\mathrm{cm^2\,s^{-1}}$. This agrees remarkably well with
observations of cloud cores by \citet{CAS02} or \citet{GOO93}. Some
prestellar cores go into collapse to build up stars and stellar
systems. The resulting protostellar objects have on average $\langle j
\rangle = 8\times10^{19}\,\mathrm{cm^2\,s^{-1}}$. This is one order of
magnitude less, and falls into the range observed in G-dwarf binaries
\citep{DUQ91b}. Collapse induced by gravoturbulent fragmentation is
accompanied by a substantial loss of specific angular momentum. This
is mostly due to gravitational torques exerted by the ambient
turbulent flow and to close encounters occurring when the protostars
are embedded in dense clusters. This eases the ``angular momentum
problem'' in star formation without invoking the presence of strong
magnetic fields.

The time evolution of $j$ is intimately connected to the mass
accretion history of a protostellar core. As interstellar turbulence
and mutual interaction in dense clusters are highly stochastic
processes, the mass growth of individual protostars is unpredictable
and can be very complex.  In addition, a collapsing cloud core can
fragment further into a binary or higher-order multiple or evolve into
a protostar with a stable accretion disk. It is the ratio of
rotational to gravitational energy~$\beta$ that determines which route
the object will take.  This is seen in the turbulent cloud cores
studied here as well as in simulations of isolated cores where
magnetic fields are important \citep[e.g.,][]{BOS99}. The
$\beta$-distribution resulting from gravoturbulent cloud fragmentation
reported here agrees well with $\beta$-values derived from
observations \citep{GOO93}. The average value is
$\beta\approx0.05$. It is important to note, that we find
that the distribution of $\beta$ stays essentially the same during
collapse and accretion \citep[see also][]{BUR00,GOO93}.

Although the accretion history and thus the evolution of the specific
angular momentum of a single protostellar object is complex, we find a
clear correlation between $j$ and mass $M$. This can be interpreted
conveniently 
assuming collapse of an initially uniform density sphere in solid body
rotation. Our models of gravoturbulent cloud fragmentation are best
represented by the relation $j\propto M^{2/3}$.

When prestellar cores form by compression as part of supersonically
turbulent flows and then go into collapse and possibly break apart
into several fragments due to the continuing perturbation by their
turbulent environment, then we expect neighboring protostars to have
similarly oriented angular momentum, at least during their early
phases of accretion.  Star clusters form hierarchically structured,
with  several young stellar objects being embedded in the same
clump of molecular cloud material.  These protostars accrete from one
common reservoir of gas and consequently gain similar specific angular
momentum. Their disks and protostellar outflows therefore will
closely align. Indeed, there are several examples of parallel disks
and outflows seen in low-mass, isolated Bok globules
\citep{FRO03,KAM03,NIS01,SAI95}. During later phases of cluster
formation, the initial substructure becomes erased by dynamical
effects and the correlation between the angular momenta of neighboring
protostars vanishes.
This is in agreement with our numerical calculations of gravoturbulent
cloud fragmentation. They show small groups of close protostellar
objects that have almost aligned specific angular momenta. As
expected, the alignment occurs during the early phase of accretion as
neighboring protostars accrete material from the same region with
similar angular momentum. During the subsequent evolution the
correlation length decreases. This is either because protostellar
aggregates disperse, or because infalling new material with different
angular momentum becomes distributed unevenly among the protostars.
  
Altogether, the process of gravoturbulent fragmentation, i.e.\ the
interplay between supersonic turbulence and self-gravity of the
interstellar gas, constitute an attractive base for a unified theory
of star formation that is able to explain and reproduce many of the
observed features in Galactic star forming regions \citep{MAC04}.
Our current study contributes with a detailed analysis of the angular momentum evolution
during collapse.

\acknowledgements{We thank Peter Bodenheimer and Mordecai Mac~Low for
  numerous stimulating discussions, and we thank our referee for
  insightful comments and suggestions. AKJ and RSK acknowledge support
  by the Emmy Noether Program of the Deutsche Forschungsgemeinschaft
  (grant no.\ KL1358/1).}

\bibliographystyle{aa}
\bibliography{myref}
\end{document}